\documentstyle[12pt,aaspp4]{article}

\begin{document}

\title{TeV Observations of Markarian 501 with the Milagrito Water Cherenkov Detector}

\author{
R.~Atkins,\altaffilmark{1}
W.~Benbow,\altaffilmark{2}
D.~Berley,\altaffilmark{3,10}
M.L.~Chen,\altaffilmark{3,11}
D.G.~Coyne,\altaffilmark{2}
R.S.~Delay,\altaffilmark{4}
B.L.~Dingus,\altaffilmark{1}
D.E.~Dorfan,\altaffilmark{2}
R.W.~Ellsworth,\altaffilmark{5}
C.~Espinoza,\altaffilmark{6}
D.~Evans,\altaffilmark{3}
A.~Falcone,\altaffilmark{7}
L.~Fleysher,\altaffilmark{8}
R.~Fleysher,\altaffilmark{8}
G.~Gisler,\altaffilmark{6}
J.A.~Goodman,\altaffilmark{3}
T.J.~Haines,\altaffilmark{6}
C.M.~Hoffman,\altaffilmark{6}
S.~Hugenberger,\altaffilmark{4}
L.A.~Kelley,\altaffilmark{2}
I.~Leonor,\altaffilmark{4}
M.~McConnell,\altaffilmark{7}
J.F.~McCullough,\altaffilmark{2}
J.E.~McEnery,\altaffilmark{1}
R.S.~Miller,\altaffilmark{6}
A.I.~Mincer,\altaffilmark{8}
M.F.~Morales,\altaffilmark{2}
M.M.~Murray,\altaffilmark{6}
P.~Nemethy,\altaffilmark{8}
J.M.~Ryan,\altaffilmark{7}
M.~Schneider,\altaffilmark{2}
B.~Shen,\altaffilmark{9}
A.~Shoup,\altaffilmark{4}
C.~Sinnis,\altaffilmark{6}
A.J.~Smith,\altaffilmark{9}
G.W.~Sullivan,\altaffilmark{3}
T.N.~Thompson,\altaffilmark{6}
T.~Tumer,\altaffilmark{9}
K.~Wang,\altaffilmark{9}
M.O.~Wascko,\altaffilmark{9}
S.~Westerhoff,\altaffilmark{2}
D.A.~Williams,\altaffilmark{2}
T.~Yang,\altaffilmark{2}
G.B.~Yodh\altaffilmark{4} \\
(The Milagro Collaboration)}

\altaffiltext{1}{University of Utah, Salt Lake City, UT\,84112, USA}
\altaffiltext{2}{University of California, Santa Cruz, CA\,95064, USA}
\altaffiltext{3}{University of Maryland, College Park, MD\,20742, USA}
\altaffiltext{4}{University of California, Irvine, CA\,92697, USA}
\altaffiltext{5}{George Mason University, Fairfax, VA\,22030, USA}
\altaffiltext{6}{Los Alamos National Laboratory, Los Alamos, NM\,87545, USA}
\altaffiltext{7}{University of New Hampshire, Durham, NH\,03824, USA}
\altaffiltext{8}{New York University, New York, NY\,10003, USA}
\altaffiltext{9}{University of California, Riverside, CA\,92521, USA}
\altaffiltext{10}{Permanent Address: National Science Foundation, Arlington, VA\,22230, USA}
\altaffiltext{11}{Now at Brookhaven National Laboratory, Upton, NY\,11973, USA}

\begin{abstract}
The Milagrito water Cherenkov detector near Los Alamos, New Mexico,
operated as a sky monitor at energies of a few TeV
between February 1997 and May 1998 including the period of the strong,
long-lasting 1997 flare of Markarian 501. Milagrito
served as a test run for the full Milagro detector.
An event excess with a significance of $3.7\,\sigma$ from Markarian 501 was observed, 
in agreement with expectations.
\end{abstract}


\section{Introduction}

Very High Energy (VHE) $\gamma$-ray astronomy studies the 
sky at energies above 100\,GeV.
To date, 4 galactic and
4 extragalactic sources have been identified as VHE sources
(see \cite{ong98} and \cite{hoffman99} for recent reviews).
The four extragalactic sources, Markarian 421
(Mrk\,421, $z=0.031$) (\cite{punch92}), 
Mrk\,501 ($z=0.034$) (\cite{quinn96}),
1ES\,2344+514 ($z=0.044$) (\cite{catanese98}) and
PKS\,2155-304 ($z=0.117$) (\cite{chadwick99}) 
are relatively nearby objects of the BL\,Lac subclass 
of active galactic nuclei. 
A characteristic feature of BL\,Lac objects is their rapid flux variability
at all wavelengths. Flaring activity 
at TeV energies has been observed both from Mrk\,421 and Mrk\,501,
ranging from variability times scales of minutes (\cite{gaidos96}) to months
(\cite{protheroe97}).

Source detections and analyses at VHE energies are currently dominated by
the highly successful atmospheric Cherenkov technique.
Cherenkov telescopes are excellent tools for the detailed study of
point sources and their sensitivity has been significantly improved over the past
few years. Strong flaring activity of Mrk\,421 and Mrk\,501 
can be detected with less than an hour of observation time per night.

To complement the pointed atmospheric Cherenkov telescopes, there is a 
strong case for wide-aperture instruments monitoring the sky with a high duty 
cycle and performing an unbiased search for new sources and source classes.
The price to pay for overcoming the limitations of atmospheric 
Cherenkov telescopes is a loss in sensitivity for 
individual sources.  To date, no unambiguous detection of a steady TeV source 
has been established with an air shower detector.

The Milagro water Cherenkov detector (\cite{mccullough99}) near
Los Alamos, New Mexico, at latitude $35.9^{\mathrm{o}}$\,N and longitude
$106.7^{\mathrm{o}}$\,W, a first-generation all-sky monitor
operating with an effective energy threshold below 1\,TeV,
started data taking in 1999.
Milagrito (\cite{atkins99}), a smaller, less sensitive prototype 
of the top layer of Milagro,
took data between February 1997 and May 1998. Milagrito was located at the same site
and served mainly as a test run for studying specific design
questions for the Milagro detector. 
Nevertheless, Milagrito operated as a fully functioning detector
and took data during the strong, long-lasting flare of Mrk\,501 in 1997.
During this flare, Mrk\,501 was intensively studied with several atmospheric 
Cherenkov telescopes (see \cite{protheroe97}).
Detailed flux and spectral studies have been published from data taken by the
Whipple telescope on Mt. Hopkins (Arizona) between February and June 1997
(\cite{samuelson98}), and the HEGRA stereo system of Cherenkov telescopes on La Palma
(Canary Islands) between March and October 1997
(\cite{aharonian99}). Although they do not cover the same
observation times, the average fluxes measured by Whipple and
HEGRA agree extremely well in both shape
and magnitude, and they both indicate an energy spectrum that deviates significantly
from a simple power law. Using an average flux as measured by Whipple,
\begin{equation}
J_{\gamma}(E) = (8.6\,\pm0.3\,\pm0.7)\times10^{-7}
		\left(\frac{E}{\mathrm{TeV}}\right)^{-2.20\,\pm0.04\,\pm0.05
		-(0.45\,\pm0.07)\,{\mathrm{log}_{10}}\,(E/\mathrm{TeV})}\,
		\mathrm{m}^{-2}\,\mathrm{s}^{-1}\,\mathrm{TeV}^{-1},
\label{equ1}
\end{equation}
simulations suggest that the observation of a statistically significant excess from Mrk\,501
is within the reach of Milagrito. 

Observations with atmospheric Cherenkov telescopes do not cover the time
between October 1997 and February 1998, when Mrk\,501 was visible only during the day time.
When atmospheric Cherenkov telescopes resumed observations in 1998, 
they observed a relatively high flux
for a few days at the beginning of March, but the flux quickly decreased and was
considerably lower for the rest of 1998 than in 1997 (\cite{quinn99}).
As an instrument that is insensitive to sunlight, Milagrito continued to monitor Mrk\,501
in late 1997 and early 1998. 
In this Letter, we present the results of an analysis of Milagrito data on Mrk\,501.

\section{The Milagrito Detector}   

As source spectra tend to be falling power laws,
a large detector area is essential for
a sufficient rate from point sources in the VHE region.
The restriction to earth-bound detectors makes the detection of the
primary $\gamma$-rays considerably more complicated,
as the primary particle generates a cascade of secondary particles in the atmosphere,
an ``air shower.'' Air shower detectors have to reconstruct the properties of the 
primary $\gamma$-ray from the secondary particles reaching the detector level, and
any $\gamma$-ray signal has to be observed in the presence of a large isotropic 
background from cosmic rays.
To achieve sufficient sensitivity at TeV energies, 
a high altitude location and the ability to detect a large
fraction of particles falling within the detector area are crucial.

Milagrito was a water Cherenkov detector of size 
$35\,\mathrm{m}\times 55\,\mathrm{m}\times 2\,\mathrm{m}$, located at 2650\,m
above sea level ($750\,{\mathrm g}\,{\mathrm cm}^{-2}$ atmospheric overburden)
in the Jemez Mountains near Los Alamos, New Mexico.
The project took advantage of an existing man-made rectangular 21 million liter
pond. A layer of 228 submerged photomultiplier tubes on a
$2.8\,\mathrm{m}\times 2.8\,\mathrm{m}$ grid detected the Cherenkov light produced 
by secondary particles
entering the water, allowing the shower direction and thus the direction
of the primary particle to be reconstructed. The detector and the detector simulation
used to study its sensitivity are described in detail elsewhere (\cite{atkins99}).

The water Cherenkov technique uses water both as the detection medium
and to transfer the energy of air shower photons to charged particles
via pair production or Compton scattering. Consequently, a large 
fraction of shower particles can be detected, leading to a high
sensitivity even for showers with primary energies below 1\,TeV.

Milagrito operated with a minimum requirement of 100 hit tubes
per event, where the discriminator threshold was
set so that a photomultiplier signal of $\sim$0.25 photoelectrons would 
fire the discriminator.
The direction of the shower plane is determined with an iterative
least squares ($\chi^{2}$) fitter using the measured times and positions
of the photomultiplier tubes. Only tubes with pulses
larger than 2 photoelectrons are used in the fit, and in subsequent
iterations, tubes with large contributions to $\chi^{2}$ are removed.
The resulting angular resolution is a strong function of the 
number of tubes remaining in the final iteration of the fit, $n_{fit}$.
If no restriction is made on $n_{fit}$, Monte Carlo simulations indicate that
the median space angle between the
fitted and the true shower direction is about $1.1^{\mathrm{o}}$ for a source
at the declination of Mrk\,501.

\begin{figure}
\plotone{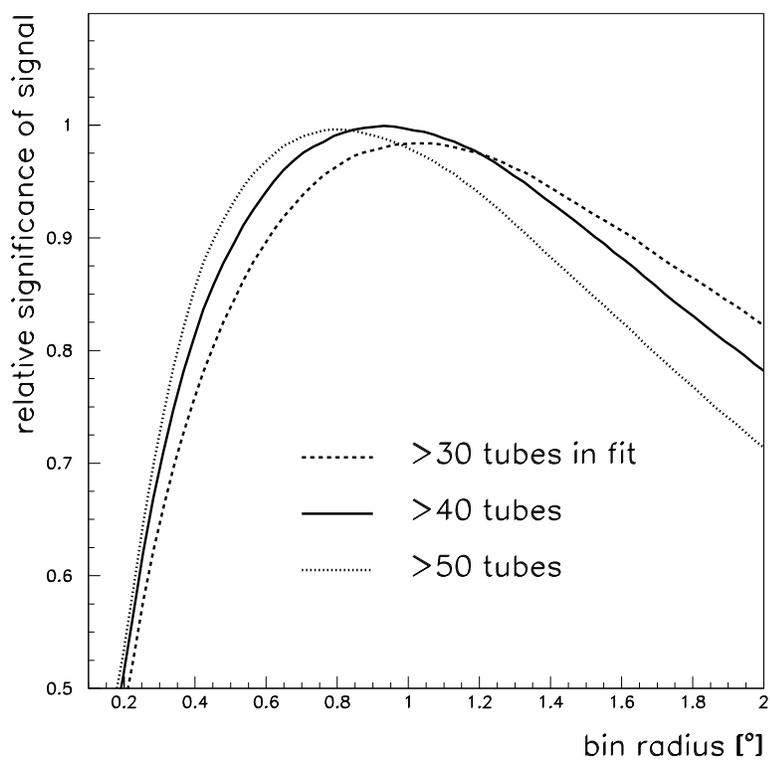}
\caption{The relative significance of a signal from a point source at the 
position of Mrk\,501 as a function of the source bin radius for various 
cuts in the number of tubes in the shower plane fit.
\label{fig1}}
\end{figure}

The optimal cut on $n_{fit}$ and the optimal 
bin size for a point source search depend upon the observed $n_{fit}$
distribution and the angular resolution as a function of $n_{fit}$.
Since the point spread function of Milagrito is not well characterized
by a two-dimensional Gaussian, the standard formulae are inappropriate.
To estimate the angular resolution as a function of $n_{fit}$,
the detector is divided into two independent, interleaved portions
(similar to a checkerboard). For each band of $n_{fit}$,
the distribution of space angle differences between the two portions
of the detector are stored. In the absence of systematic effects, 
these distributions can be interpreted
as twice the point spread function of the detector for the given band
of $n_{fit}$ (\cite{alexandreas92}). Under the assumption that the 
point spread function for $\gamma$-ray showers is identical 
to that of hadron-induced air showers, one can use the above distributions
to determine the optimal cut on $n_{fit}$ and the optimal size of the
angular bin.
Figure\,\ref{fig1} shows the expected significance of a source
as a function of angular bin size for three different cuts on $n_{fit}$.
The analysis indicates that requiring $n_{fit}>40$ with
a bin size of radius $1.0^{\mathrm{o}}$, which on average contains
$57\,\%$ of the source events, is optimal for a binned analysis.
As shown in Figure\,\ref{fig1}, for a rather wide range of cuts,
the significance of an excess depends only weakly on the chosen source bin size.

As the detector is much smaller than the typical lateral size of a shower,
the shower core, {\it i.e.} the point where the primary 
particle would have struck the detector had there been no atmosphere,
is outside the sensitive detector area 
for a large fraction of showers fulfilling the trigger condition.
Assuming a differential flux following $E^{-2.8}$ for the proton 
background and $E^{-2.5}$ for a typical $\gamma$-source,
$16\,\%$ of the proton showers and $21\,\%$ of the $\gamma$-showers
triggering the Milagrito detector have their cores on the pond.
This leads to a broad distribution of detected events 
with no well defined threshold energy. 
Monte Carlo simulations using the Mrk\,501 spectrum given in 
Equation\,\ref{equ1} predict a distribution starting at energies
as low as 100\,GeV, with $90\,\%$ of the detected events 
having an energy in excess of 0.8\,TeV.
The median energy of detected showers depends
on the declination $\delta$ and the spectral index of the source, and typical values
are 3\,TeV for $\delta=39.8^{\mathrm{o}}$ (Mrk\,501) and
7\,TeV for $\delta=22.0^{\mathrm{o}}$ (Crab nebula, assuming an $E^{-2.5}$
spectrum).

Detector performance is best evaluated by observations of well-known
sources. The standard candle of VHE astronomy is the Crab nebula.
Simulations indicate that the expected statistical significance 
of the excess above background from
the Crab nebula in Milagrito is too small to be used for testing
Milagrito's performance, and indeed no significant excess from this source
was observed.
However, the large average flux of Mrk\,501 during its flaring state in 1997
results in an expected event rate from Mrk\,501
3.6 times the Crab rate for Milagrito.
Mrk\,501 can therefore be used to measure the sensitivity of Milagrito and
to test the reliability of the detector simulation.

\section{Results}

Milagrito took data on Mrk\,501 from February 8, 1997 to May 7, 1998.
The effective exposure time was about 370 days, with most of the downtime being
due to power outages, detector maintenance, and upgrades.
Milagrito started operation with about 0.9\,m of water above the tubes.
The water level was increased starting in November 1997
to study how the sensitivity changes with water depth. 
The trigger rate was about 300\,Hz with 0.9\,m of water and increased to
340\,Hz (400\,Hz) at a depth of 1.5\,m (2.0\,m).

The measured rate of $2420\pm 80$ reconstructed events per day 
for 0.9\,m water depth in a typical bin with $1.0^{\mathrm{o}}$ radius 
at the same declination as Mrk\,501 is in good agreement
with the predicted rate of $2460^{+160}_{-90}$ events per day from 
protons, Helium, and CNO nuclei
(the error accounts for the uncertainty in the measured flux). 
The contributions from He and CNO 
to this predicted rate are $27\,\%$ and $4\,\%$, respectively.

The isotropic cosmic ray background flux exceeds the 
$\gamma$-signal from Mrk\,501 by several orders of magnitude.  The 
expected background flux in the source bin must be subtracted from the
measured one in order to obtain the number of excess events from the
source. Since the background in the source bin depends on its exposure 
and the detector efficiency in local angular coordinates, the background is
calculated directly from the data (\cite{alexandreas93}).
For each detected event, ``fake'' events are generated by keeping the local zenith 
and azimuth angles ($\theta, \phi$) fixed and calculating new values for right
ascension using the times of 30 events randomly selected from a buffer 
that spans about 2 hours of data taking. The background level is then calculated
from the number of fake events falling into the source bin.
By using at least 10 fake events per real event, the
statistical error on the background can be kept sufficiently small.

\begin{figure}
\plotone{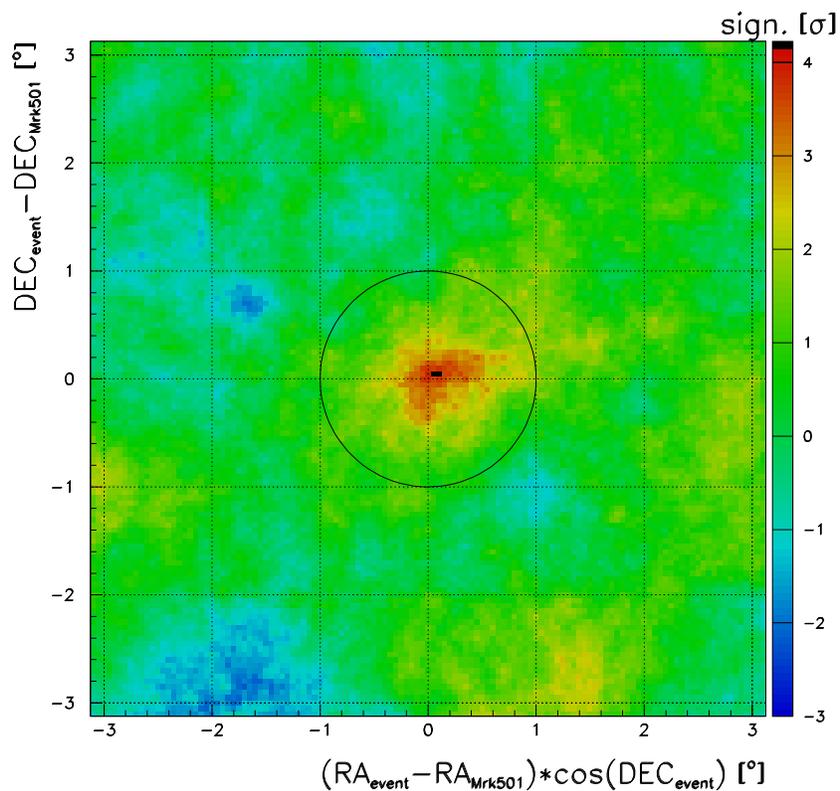}
\caption{
The significance for an event excess as a function of right ascension (RA)
and declination (DEC) in a $6^{\mathrm{o}}\times 6^{\mathrm{o}}$ region with the Mrk\,501
position (RA=$253.468^{\mathrm{o}}$, DEC=$39.760^{\mathrm{o}}$ (J2000)) 
in the center.
For each bin, the significance is calculated for the area of
the circle with radius $1.0^{\mathrm{o}}$ and the bin center as the central point,
thus neighboring bins are highly correlated. The circle indicates the source bin.
\label{fig2}}
\end{figure}

Figure\,\ref{fig2} shows the significance of the observed signal as 
a function of right ascension
and declination in a $6^{\mathrm{o}}\times 6^{\mathrm{o}}$ region with the Mrk\,501
position in the center. For each bin, the significance is calculated for the area of
the circle with radius $1.0^{\mathrm{o}}$ and the bin center as the central point,
hence neighboring bins are highly correlated.

At the source position, 918\,954 events are
observed with an average expected background of $915\,330\pm 250$ events.
The excess of $3624\pm 990$ events corresponds to a significance of 3.7 sigma.
We interpret this result as a reconfirmation of Mrk\,501 as a TeV 
$\gamma$-ray source during this period.
The corresponding excess rate averaged over the lifetime of Milagrito
is $(9.8\pm 2.7)\,\mathrm{day}^{-1}$. 
The excess rate measured between February and
October 1997 can be directly compared to the $\gamma$-rate expected using
the average flux measured by atmospheric Cherenkov telescopes during this period.
Using the flux given in Equation\,\ref{equ1}, Monte Carlo
simulations of a full source transit predict a $\gamma$-rate of
$(12.5\pm 3.8)\,\mathrm{day}^{-1}$, which is in good agreement
with the measured rate during this period of $(13.1\pm 4.0)\,\mathrm{day}^{-1}$.

\begin{figure}
\plotone{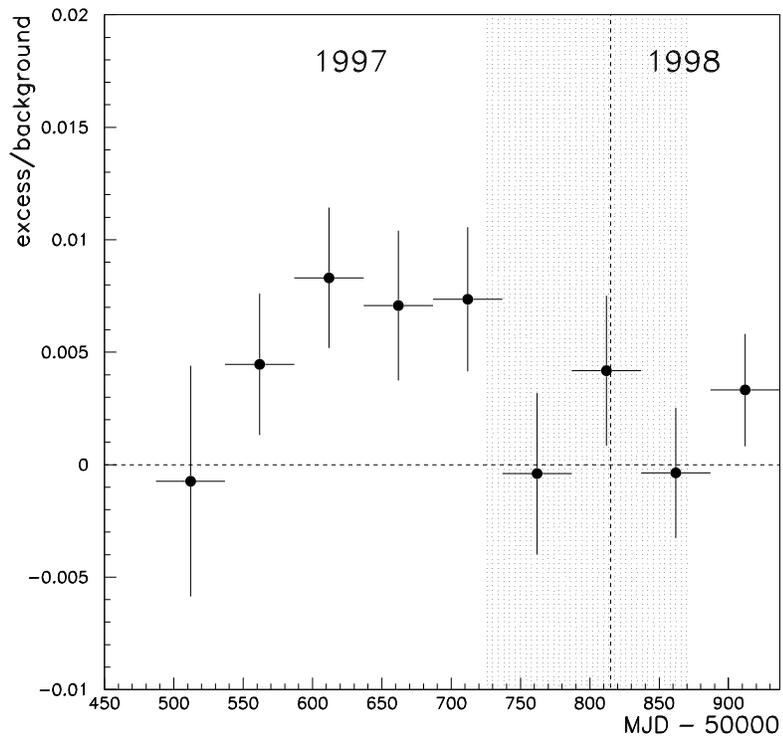}
\caption{
The fractional event excess from Mrk\,501 as a function of time. 
The shaded area indicates the time period for which no data
from atmospheric Cherenkov telescopes is available.
\label{fig3}}
\end{figure}

Figure\,\ref{fig3} shows excess divided by background for the lifetime of Milagrito.
At Milagrito's level of sensitivity, the flux is consistent with being
constant in time.

The analysis was extended to 10 other nearby blazars ($z<0.06$) in
Milagrito's field of view, including Mrk\,421, but Mrk\,501 remains the
only analyzed source with a significance in excess of 3\,$\sigma$. Results
from this blazar sample are reported elsewhere (\cite{westerhoff98}).

\section{Conclusions and Outlook}

Milagrito, the first TeV air shower detector based on the water Cherenkov
technique, observed an excess with a statistical significance of $3.7\,\sigma$
from the direction of Mrk\,501 between February 1997 and May 1998.
The excess is in agreement with expectations based on simulations and indicates
that the technique is working as anticipated.

Milagrito served as a prototype for the full Milagro detector. In its final stage, 
Milagro has a size of $60\,\mathrm{m}\times 80\,\mathrm{m}\times 8\,\mathrm{m}$
and two layers of photomultiplier tubes, an upper layer with 450 tubes at a depth
of 1.5\,m, and an additional layer with 273 tubes at a depth of 6.2\,m. 
With its larger effective area and the ability to reject some of the cosmic ray background,
Milagro will be at least 5 times as sensitive as Milagrito.
Data taking began in early 1999.

\acknowledgments

This research was supported in part by the National Science Foundation, 
the U.S. Department of Energy Office of High Energy Physics, 
the U.S. Department of Energy Office of Nuclear Physics, 
Los Alamos National Laboratory, the University 
of California, the Institute of Geophysics and Planetary Physics, 
The Research Corporation, and the California Space Institute.

\clearpage

\clearpage

\clearpage

\end{document}